\documentclass[aps,prb,10pt,citeautoscript,reprint]{revtex4-1}

\usepackage[hidelinks,colorlinks=true,allcolors=blue]{hyperref}
\usepackage{amssymb}
\usepackage{graphicx}
\usepackage{physics}
\usepackage{dcolumn}
\usepackage{multirow}
\usepackage{bm}
\newcolumntype{.}{D{.}{.}{1}}

\graphicspath{ {./}{../figures/} }

\newcommand{\rr}{\mathbf{r}}
\newcommand{\RR}{\mathbf{R}}
\newcommand{\bb}{\mathbf{b}}
\newcommand{\kk}{\mathbf{k}}
\newcommand{\nk}{n\mathbf{k}}
\newcommand{\OmegaI}{\Omega_{\text{I}}}
\newcommand{\OmegaD}{\Omega_{\text{D}}}
\newcommand{\OmegaOD}{\Omega_{\text{OD}}}
\newcommand{\OmegaIOD}{\Omega_{\text{I},\text{OD}}}

\newcolumntype{d}{D{.}{.}{4}}

\begin{document}

\title{
Automated construction of maximally localized Wannier functions:
Optimized projection functions method
}
\author{Jamal I. Mustafa}
\email{jimustafa@berkeley.edu}
\author{Sinisa Coh}
\author{Marvin L. Cohen}
\author{Steven G. Louie}
\affiliation{Department of Physics, University of California at
  Berkeley and Materials Sciences Division, Lawrence Berkeley National
  Laboratory, Berkeley, California 94720, USA}

\begin{abstract}
Maximally localized Wannier functions are widely used in electronic
structure theory for analyses of bonding,
electric polarization, orbital magnetization, and for interpolation. The state of
the art method for their construction is based on the
method of Marzari and Vanderbilt.
One of the practical difficulties of this method is guessing 
functions (initial projections) that approximate the final Wannier functions. 
Here we present an approach based on optimized projection functions
that can construct maximally localized Wannier functions without a guess.
We describe and demonstrate this approach on several realistic examples.
\end{abstract}

\maketitle

\section{Introduction and motivation}\label{sec:intro}

Within the quasiparticle approximation, the electronic states of a
crystal can be described in terms of single-particle Bloch functions $\psi_{\nk}(\rr)$. These
functions are eigenstates of the crystal Hamiltonian, and can be labeled by
their band index $n$ and crystal momentum $\kk$.
Wannier functions (WFs) provide an alternative representation in which an
entire band of electrons is described by a single function $\ket{\RR n}$
localized in or near the unit cell labeled by the lattice vector $\RR$.
In their simplest form, WFs are obtained from the Bloch
functions via the Fourier transformation
\begin{equation}\label{eq:wann}
  \ket{\RR n}=\frac{V}{{(2\pi)}^{3}}\int_{\textrm{BZ}}\textrm{d}\mathbf{k}\,e^{-i\kk\cdot\RR}\ket{\psi_{\nk}},
\end{equation}
where $V$ is the volume of the real-space primitive cell.
The definition of WFs is not unique because there is a gauge freedom in the
right-hand side of Eq.~\eqref{eq:wann}.  Namely, at each $\kk$ point and
for each $n$, one can change the overall phase of the Bloch state
$\ket{\psi_{\nk}}$. In fact, one often considers an even more general
gauge choice which allows an arbitrary unitary transformation of a set
of $N$ bands at each $\kk$ point,
\begin{equation}\label{eq:gauge}
  \ket{\psi_{n\kk}} \rightarrow \sum_{m} u^{(\kk)}_{m n}
  \ket{\psi_{m\kk}}.
\end{equation}
We focus here on the case when these $N$ bands are isolated from the 
rest.
The choice of gauge is now expressed through a $\kk$-dependent $N\!\times\!N$
unitary matrix $u^{(\kk)}$.

When Wannier functions are localized in real space they have a wide use 
in the electronic structure community.  An extensive review of maximally
localized Wannier functions (MLWFs) and their properties and applications can be found in
Ref.~\citenum{RevModPhys.84.1419}. For example, they have been 
used in the description of electronic
polarization\cite{PhysRevLett.97.107602} and orbital magnetization, in addition
to being used for
interpolation of band structures and matrix elements
\cite{PhysRevB.75.195121,PhysRevB.76.165108,Noffsinger20102140} and
electron transport calculations.\cite{Pizzi2014422} 

For this reason, one often uses the gauge freedom $u^{(\kk)}$ so that the 
corresponding WFs are localized.
As a general consequence of the Fourier transform, the localization of
the WFs $\ket{\RR n}$ in $\rr$ space will depend on the smoothness of the
gauge $u^{(\kk)}$ in $\kk$ space.
If the $\psi_{\nk}(\rr)$ are chosen with
random overall complex phases [which often happens if
$\psi_{\nk}(\rr)$ are acquired numerically by diagonalizing a $\kk$-dependent Hamiltonian matrix separately for each $\kk$ point] the WFs obtained from
Eq.~\eqref{eq:wann} need not be localized.  However, if matrices
$u^{(\kk)}$ are chosen so that Bloch states are smooth in
$\kk$ space (smooth gauge), the corresponding WFs will be localized in
$\rr$ space.

The idea of maximally localized Wannier functions and a procedure for
obtaining them from a set of composite bands was introduced by Marzari and
Vanderbilt\cite{PhysRevB.56.12847} for isolated bands
and later extended to the case of
entangled bands.\cite{PhysRevB.65.035109} Maximally localized Wannier functions
are constructed by choosing a gauge $u^{(\kk)}_{mn}$ for Eq.~\eqref{eq:gauge}
that minimizes the spread functional
\begin{align}
  \Omega           &= \sum_{n}\qty[\expval{r^2}_n-\overline{\rr}_n^2] \label{eq:omega},
\end{align}
where
\begin{align}
  \expval{r^2}_n   &= \expval{r^2}{\mathbf{0}n},                      \label{eq:r2n}    \\
  \overline{\rr}_n &= \expval{\rr}{\mathbf{0}n}.                       \label{eq:rn}
\end{align}
Here the spread functional $\Omega$ is written in terms of the Wannier
functions $\ket{\mathbf{0}n}$.
Usually there exists a global minimum of $\Omega$ corresponding to a unique
choice of $u^{(\kk)}$ (up to translation of the WFs and their overall complex phase), but in some cases there
are multiple solutions.\cite{PhysRevB.56.12847}

Using the general form of Eq.~\eqref{eq:wann} including the $u^{(\kk)}$ matrix in Eq.~\eqref{eq:gauge}, the spread can be recast in terms of the
Bloch states. More specifically, $\Omega$ can be expressed only as a
function of the overlaps of the periodic parts of the Bloch functions at
neighboring $\kk$ points $\kk$ and $\kk+\mathbf{b}$,
\begin{align}
  m^{(\kk,\bb)}_{ij} = \braket{u_{i\kk}}{u_{j\kk+\mathbf{b}}} \label{eq:def_m}.
\end{align}
See Appendix~\ref{sec:appndx-spread} for an explicit definition of $\Omega$
in terms of $m^{(\kk,\bb)}$,
\begin{align}
\Omega = \Omega\qty[m^{(\kk,\bb)}].
\end{align}
Here we only note that spread $\Omega$ can be decomposed into three
parts: the invariant part, which does not depend on the gauge, and the
diagonal part and the off-diagonal parts which do,
\begin{align}
  \Omega\qty[m^{(\kk,\bb)}]=\OmegaI+\OmegaD+\OmegaOD.
 \label{eq:three_parts}
\end{align}

The procedure for minimizing $\Omega$, outlined in
Refs.~\citenum{PhysRevB.56.12847} and \citenum{PhysRevB.65.035109}, is
implemented in the Wannier90 code\cite{Mostofi2008685} and has become
the standard method for obtaining localized WFs.
A notable drawback in the standard approach that we address in this 
manuscript is that one often needs to provide a good initial guess of the
MLWFs to find the global minimum of $\Omega$.
In this work we demonstrate a modified procedure, in which
localized Wannier functions are constructed as a linear combination of physically based atom-centered orbitals
without requiring an initial guess, as in the standard 
approach.\cite{PhysRevB.56.12847}
This is achieved by finding optimal projection functions (OPFs) so that the
resulting Wannier functions obtained via projection (as in Sec.~\ref{sec:step1})
are as localized as possible.
This OPF method could, for example, be used in constructing 
material properties databases,
such as the database of the \textit{Materials Project},\cite{Jain2013} by providing a simple
localized Hamiltonian that could serve as a descriptor for
the electronic structure of a material.
We present the theoretical approach and numerical methods in Sec.~\ref{sec:stdsol}~and~\ref{sec:altsol}.
Several realistic materials are investigated in Secs.~\ref{sec:examples} to 
illustrate our approach for constructing localized WFs.

Schemes beyond the standard
implementation\cite{PhysRevB.56.12847,PhysRevB.65.035109,Mostofi2008685} have
been developed by others to improve the construction of MLWFs and their properties. The
inclusion of unoccupied anti-bonding states has been
shown\cite{PhysRevLett.94.026405,*PhysRevB.72.125119} to give more localized
Wannier functions, but at the expense of a chemical picture of the occupied
states.
Additionally, constraints on the $u^{(\kk)}$ matrices can be imposed
in order to construct localized Wannier functions that possess all the
space group
symmetries of the crystal.\cite{PhysRevB.87.235109}

\section{Standard approach}\label{sec:stdsol}

Here we summarize the main result of
Ref.~\onlinecite{PhysRevB.56.12847} for a two-step construction of
maximally localized Wannier functions.
In the first step of minimizing the spread
functional $\Omega$ 
one needs to
guess orbitals $g_j(\rr)$ with
roughly the same orbital characters and real-space location $\overline{\rr}_j$ as the
target MLWFs.  This choice is often done based on an intuitive
understanding of the band structure of the crystal under
investigation.  Given a choice of $g_j(\rr)$ close to target WFs, one
constructs the gauge for which spread functional $\Omega$ is {\it
near} its global minimum (better choices give $\Omega$ closer to the
global minimum).  In the second step, this initial gauge choice is
iteratively optimized until $\Omega$ reaches a global minimum.
In practice, the second step usually reduces the spread $\Omega$ only by 20\%
or less.

\subsection{First step}\label{sec:step1}

Now we describe the first step of this procedure in the
simple case of a single band of states $\psi_{\kk}(\rr)$.  Given a
localized function $g(\rr)$ approximating the target MLWF at the origin, we first project
it onto the Bloch state $\psi_{\kk}(\rr)$ at each $\kk$
\begin{equation}\label{eq:single-band_a}
  a^{(\kk)}=\braket{\psi_{\kk}}{g}.
\end{equation}
Now we rotate the phase of Bloch state $\psi_{\kk}(\rr)$ so that the
relative phase of rotated Bloch state and $g(\rr)$ is zero for all
$\kk$ points,
\begin{equation}\label{eq:single-band_psi}
  \ket{\psi_{\kk}} \rightarrow
  a^{(\kk)}\qty(a^{(\kk)*}a^{(\kk)})^{-1/2}\ket{\psi_{\kk}}.
\end{equation}
It is easy to check that if the initial guess $g(\rr)$ were a true target MLWF,
inserting these rotated Bloch states into Eq.~\eqref{eq:wann} would give
back the target MLWF.
However, since $g(\rr)$ is only an approximation,
the spread $\Omega$ of the rotated Bloch states is not exactly at the
global minimum. For a good guess $g(\rr)$, however, the spread should be close
to the global minimum.

Following the procedure in the case of a single band, we now
generalize it to the case of $N$ composite bands.  First, we choose a
set of $N$ localized orbitals $g_j(\rr)$ that are approximately equal to
the $N$ target MLWFs,
\begin{align}
\ket{g_j} \approx \ket{\mathbf{0} j}.
\label{eq:condition1}
\end{align}
Here we choose for convenience $\ket{g_j}$ to be close to the MLWFs near the
origin ($\RR=\mathbf{0}$),
but in principle any other $\RR$ can be chosen.

Next we compute the overlap between all $N$ Bloch bands
and $N$ initial guesses for the WFs,
\begin{equation}\label{eq:proj-multi}
  a^{(\kk)}_{ij}=\braket{\psi_{i\kk}}{g_j}.
\end{equation}
Unlike the case of a single band, $a^{(\kk)}$ for an isolated group of N Bloch bands is a $N\!\times\!N$
matrix, so that Eq.~\eqref{eq:single-band_psi} generalizes to
\begin{equation}
  \ket{\psi_{i\kk}} \rightarrow \sum_{j} u^{(\kk)}_{ji}
  \ket{\psi_{j\kk}} = \sum_{jl} a_{jl}
  \qty[\qty(a^{\dagger}a)^{-1/2}]_{li} \ket{\psi_{j\kk}}.
  \label{eq:gen_mb}
\end{equation}
Here, to simplify notation, we suppress the $\kk$ dependence of $a$.
The inverse square root on the right-hand side is the
matrix square root of $\qty(a^{\dagger}a)^{-1}$.
For further simplification, define $u_x$, for an arbitrary matrix $x$, as
\begin{equation}\label{eq:uaN}
  u_x \equiv x ( x^{\dagger} x )^{-1/2}.
\end{equation}
The matrix $u_x$ is unitary by construction.  In fact, it is the closest
unitary approximation of $x$.

In practice, the unitary matrix $u_x$ is obtained via the singular value
decomposition (SVD) of $x$.  If $x=zdv$ is a SVD with $z$ and $v$ unitary,
and $d$ diagonal, then $u_x$ is simply $zv$.
Given this notation, the earlier gauge transformation from Eq.~\eqref{eq:gen_mb}
now reads
\begin{equation}
  \ket{\psi_{i\kk}} \rightarrow \sum_{j} \qty[u^{(\kk)}_a]_{ji}
 \ket{\psi_{j\kk}}.
\label{eq:rotatedN}
\end{equation}
As in the case of a single band, if trial orbitals $g_j(\rr)$
were chosen close enough to the target MLWF, the gauge $u^{(\kk)}_a$ from
Eq.~\eqref{eq:rotatedN} will by definition give a spread
close to the global minimum,
\begin{equation}\label{eq:first_step_formal}
 \Omega=\Omega\qty[ u^{(\kk)\dagger}_{a}m^{(\kk,\bb)}u^{(\kk+\bb)}_{a} ].
\end{equation}
Here we implicitly wrote $\Omega$ in terms of
overlap matrices $m^{(\kk,\bb)}$ and used gauge transformation of
Bloch states from Eq.~\eqref{eq:gauge} to get transformation of the overlap
matrix $m^{(\kk,\bb)}$ defined in Eq.~\eqref{eq:def_m}.

\subsection{Second step}

The initial gauge $u^{(\kk)}_a$ can be further improved in the second
step by rotating, at each $\kk$ point, the gauge from $u^{(\kk)}_a$ to
$u^{(\kk)}_a v^{(\kk)}$
with an appropriate choice of $\kk$-dependent matrices $v^{(\kk)}$.
The spread functional $\Omega$ is minimized using the method of steepest
descent. The gradient is determined by calculating the derivative of the spread
with respect to the unitary matrices $v^{(\kk)}$ and then following the path along
the direction which minimizes $\Omega$.
Written more formally, the second step of the standard procedure finds
a set of unitary $N\!\times\!N$ matrices
\begin{equation}\label{eq:stdsol-opt1}
 v^{(\kk)} \in \mathcal{U}(N,N), \text{ one for each $\kk$,}
\end{equation}
that
\begin{equation}\label{eq:stdsol-opt2}
\text{minimize} \quad \Omega\qty[v^{(\kk)\dagger}u^{(\kk)\dagger}_{a}m^{(\kk,\bb)}u^{(\kk+\bb)}_{a}v^{(\kk+\bb)}].
\end{equation}

Quite generally, the global minimization of a function using the
steepest descent algorithm is bound to work well when one starts near
the global minimum.  Otherwise it is quite possible for the algorithm
to get stuck in a local minimum.  In other words, the second step of the
procedure will arrive at the true MLWFs as long as the initial guesses $g_j(\rr)$ in
the first step are close enough.
It is this issue that we aim to address in this manuscript: how to
automatically construct a gauge that is guaranteed to be close to the
global minimum.

\section{Alternative approach}\label{sec:altsol}

In our approach, instead of choosing $N$ functions $g_j(\rr)$ that are
close to the $N$ target MLWFs, we start with a larger set of $M$ functions
($M\!\geq\!N$) labeled $h_j(\rr)$.  These functions $h_j$ will be chosen
so that any MLWF near the origin ($\RR=\mathbf{0}$) can approximately be written as a
linear combination of $h_j$.  In other words, the space spanned by $h_j$ must 
approximately contain, as a subset, the space spanned by the MLWFs near the origin,
\begin{align}
\textrm{Span} \qty( \ket{h_j} )
\supseteq
\textrm{Span} \qty( \ket{\mathbf{0} n} ).
\label{eq:condition2}
\end{align}
The requirement on $\ket{h_j}$
is significantly less restrictive than that
on $\ket{g_j}$ in the standard approach.
In fact, the requirement Eq.~\eqref{eq:condition2} should be rather easily satisfied.
Since we expect MLWFs to be linear combinations of
atomiclike valence electrons, we can simply choose $h_j$ to be a set of atom-centered
atomic orbitals for each atom in the crystal basis and for each
relevant atomiclike orbital in the valence (some combination of
$s$, $p$, $d$, $f$ atomic orbitals, depending on the valence).  
If nominal valence atomiclike orbitals
are not enough to satisfy Eq.~\eqref{eq:condition2} (which might happen for example
in material under extreme pressure), one can always include atomic orbitals
with higher radial and orbital quantum numbers.

In the case of covalently
bonded materials, a specific target MLWF might have its center on a covalent bond at
the edges of the primitive unit cell. If this is the case, then we
can expand the set $h_j(\rr)$ by including the periodic images of a
few atoms in the crystal basis, so that in the end,
for each unique covalent bond, both atoms forming the bond are
included in $h_j(\rr)$.

Since the functions $h_j$ satisfy Eq.~\eqref{eq:condition2}, it is
possible to approximate the MLWFs as linear combinations of
$h_j$. Formally, it is possible to find a semiunitary rectangular
$M\!\times\!N$
matrix $W$ such that the functions
\begin{equation}
  \ket{\bar{g}_j} = \sum_{i=1}^M W_{ij}\ket{h_i}
\end{equation}
are close to the target MLWFs.
(Since $W$ is rectangular, the $N$ functions $\bar{g}_j$ are linear combinations of $M$ functions $h_j$.)
Thus obtaining approximate MLWFs is equivalent to finding the matrix $W$.
We shall call these $\bar{g}_j$ optimized projection functions (OPFs).

To measure the closeness of $\bar{g}_j$ to the target MLWFs, we need to
express spread $\Omega$ in terms of $W$.  Therefore, we first need a
projection of $\bar{g}_j$ into Bloch states.  Since $\bar{g}$ depends
on $W$, it is more convenient to first project $h_j$ onto the Bloch
states, yielding the $N\!\times\!M$ projection matrix
\begin{equation}
  A^{(\kk)}_{ij}=\braket{\psi_{i\kk}}{h_{j}}. \label{eq:expanded_A}
\end{equation}
Given $A^{(\kk)}$ we can compute the overlap matrix between the $\bar{g}_j$
and the Bloch states,
\begin{equation}
  \bar{a}^{(\kk)}_{ij}
  = \braket{\psi_{i\kk}}{\bar{g}_j}
  = \sum_{l=1}^M \braket{\psi_{i\kk}}{h_l}W_{lj},
\end{equation}
or, in short,
\begin{equation}
\bar{a}^{(\kk)}=A^{(\kk)}W. \label{eq:aaw}
\end{equation}
Here we adopted the convention that small ($N\!\times\!N$) square
matrices are written with lower-case Latin letters, while rectangular
($N\!\times\!M$ or $M\!\times\!N$) or large square matrices ($M\!\times\!M$)
are denoted by upper-case Latin letters.

Now we are ready to express $\Omega$ in terms of $W$. Combining
Eq.~\eqref{eq:first_step_formal} and Eq.~\eqref{eq:aaw} yields
\begin{equation}
  \Omega=\Omega\qty[u^{(\kk)\dagger}_{AW}m^{(\kk,\bb)}u^{(\kk+\bb)}_{AW}].
\end{equation}
To draw comparison with Eqs.~\eqref{eq:stdsol-opt1}
and~\eqref{eq:stdsol-opt2}, in our approach the process of constructing
MLWFs is equivalent to finding
\begin{equation}\label{eq:altsol-opt1}
   W \in \mathcal{U}(M,N), \text{ a single matrix}
\end{equation}
that
\begin{equation}\label{eq:altsol-opt2}
\text{minimizes} \quad \Omega\qty[u^{(\kk)\dagger}_{AW}m^{(\kk,\bb)}u^{(\kk+\bb)}_{AW}].
\end{equation}
Once the $W$ which minimizes Eq.~\eqref{eq:altsol-opt2} is found, we use the
matrices $u^{(\kk)}_{AW}$ to rotate Bloch states at each $\kk$ point into a 
smooth gauge.  In most of the concrete cases studied, the spread of the Wannier functions
corresponding to this gauge is within 1\% of the global minimum (this is discussed 
further in Sec.~\ref{sec:examples}) and therefore there is no need to improve
the gauge further.  However, in principle one could run the second step of the standard 
procedure to bring spread to its true global minimum and thus obtain {\it maximally}
localized Wannier functions.

Now we will compare our approach to the standard method in more
detail, outlining both the advantages and disadvantages of our
approach. We also discuss the approximations that are made to
implement an algorithm to construct the $W$ matrix.

\subsection{Comparison to the standard approach}

The procedure for constructing MLWFs by generating OPFs
[Eqs.~\eqref{eq:altsol-opt1} and~\eqref{eq:altsol-opt2}] has several
advantages compared to the standard procedure
[Eqs.~\eqref{eq:stdsol-opt1} and ~\eqref{eq:stdsol-opt2}]. First, OPF
construction is given by a single matrix $W$, instead of a set of
$v^{(\kk)}$ matrices, one at each $\kk$ point.
For this reason, as will be shown in Sec.~\ref{sec:approx}, one can
more directly solve Eq.~\eqref{eq:altsol-opt2} without using the
method of steepest descent; rather, an iterative procedure is used to
construct $W$ as a product of large unitary transformations 
(Givens rotations).
Therefore, this procedure is less likely to get stuck in a local minimum.
The second advantage of OPF construction is that the $W$ matrix itself has
a lot of chemical information
encoded in it.  For example, one can see directly from $W$ the
contribution of the various atomic orbitals to each OPF and thus the
corresponding Wannier functions. We discuss this point on concrete
examples in Sec.~\ref{sec:examples}.
Third, the use of a single matrix might make it easier to impose
constraints such as crystal symmetry.

There are however some disadvantages to the OPF construction approach.
First, the spread $\Omega$ in Eq.~\eqref{eq:altsol-opt2} depends
nonlinearly on $W$ since it appears under the matrix
inverse square root in $u^{(\kk)}_{AW}$.  In fact, Taylor
expansion of the inverse square root leads to a power series in all
positive integer powers of $W$.  Second, since we do not want to rely on a
steepest decent method, minimization of the diagonal part of the spread
[$\OmegaD$ in Eq.~\eqref{eq:three_parts}] becomes nontrivial.

In the following section we introduce two simplifications to
Eq.~\eqref{eq:altsol-opt2} which deal with these two disadvantages of
OPF and allow for an efficient numerical construction of OPFs in all the
cases studied.

\subsection{Simplifications}\label{sec:approx}

The following two subsections describe two simplifications that turn
minimization of Eq.~\eqref{eq:altsol-opt2} into a numerically
efficient form.

\subsubsection{Linearizing \texorpdfstring{$u_{AW}$}{u\_AW}}\label{subsec:approx-linearize}

The first simplification in minimizing the spread $\Omega$ from
Eq.~\eqref{eq:altsol-opt2} is to expand it to the leading order in
$W$. Explicitly writing $u_{AW}$ in terms of its definition
[Eq.~\eqref{eq:uaN}] and ignoring $\kk$ index for the moment,
\begin{align}
u_{AW} = AW \left( W^{\dagger} A^{\dagger} A W \right)^{-1/2}.
\end{align}
For $W$ which minimizes Eq.~\eqref{eq:altsol-opt2} we expect
\begin{align}
W^{\dagger}A^{(\kk)\dagger}A^{(\kk)} W \approx I_N,
\label{eq:condition_waaw}
\end{align}
for all $\kk$ since the OPFs approximately overspan the space of MLWFs ($I_N$ is
the $N\!\times\!N$ identity matrix). Therefore, at least near the
optimal value of $W$, we are justified in Taylor
expanding $u_{AW}$ around $W^{\dagger}A^{\dagger}AW$ close to the
identity ($I_{N}$),
\begin{align}
u_{AW} = AW \qty[ I_N - \frac{1}{2} \qty(W^{\dagger}A^{\dagger}AW - I_N) + \ldots].
\end{align}
Therefore, to lowest order, $u_{AW} \approx AW$. Restoring unitarity we
can replace $A$ with $U_A$, thus obtaining a unitary approximation to
$U_{AW}$,
\begin{align}
u_{AW} \approx U_{A}W.
\label{eq:approx_uaw}
\end{align}
Here $U_A$ has been constructed according to the L\"owdin
orthonormalization procedure given by Eq.~\eqref{eq:uaN}.  We
follow here the notation we introduced earlier so that $U_A$ with
upper case $U$ is a rectangular $N\!\times\!M$ matrix (while $u_{AW}$
with lowercase $u$ is a square $N\!\times\!N$ matrix).
We also note here that Eq.~\eqref{eq:approx_uaw} is exact if $W$ were a square
matrix.

Inserting Eq.~\eqref{eq:approx_uaw} into Eq.~\eqref{eq:altsol-opt2} we
find that construction of OPFs is equivalent to finding a rectangular matrix
$W \in \mathcal{U}(M,N)$ that
\begin{equation}
\text{minimizes} \quad \Omega\qty[W^{\dagger}U^{(\kk)\dagger}_{A}m^{(\kk,\bb)}U^{(\kk+\bb)}_{A}W].
\label{eq:approx_condition}
\end{equation}
Here, $U^{\dagger}_{A}mU_{A}$ is identified as the enlarged
($M\!\times\!M$) overlap matrices projected into the space of $M$ orbitals
$h_j$.

In most cases, the $W$ that minimizes
Eq.~\eqref{eq:approx_condition} also satisfies
Eq.~\eqref{eq:condition_waaw}, which then justifies the Taylor
expansion of $u_{AW}$.  However, occasionally this is not
the case (for example, in strongly covalent materials with a lot
of symmetry).
Therefore, we will introduce a Lagrange multiplier $\lambda$ to
Eq.~\eqref{eq:approx_condition}, which imposes condition
Eq.~\eqref{eq:condition_waaw}.  With this modification, we now seek
matrix $W$ and $\lambda$ at a saddle point of the Lagrangian,
\begin{align}
\mathcal{L} \qty(W,\lambda) = &
\Omega\qty[W^{\dagger}U^{(\kk)\dagger}_{A}m^{(\kk,\bb)}U^{(\kk+\bb)}_{A}W]
\notag \\
+ & \lambda w \sum_{\kk}\sum_{i=1}^N \abs{ \qty[W^{\dagger}A^{(\kk)\dagger}A^{(\kk)}
W]_{ii} - 1 }^2.
\label{eq:with_lagrange}
\end{align}
For convenience 
we rescaled the Lagrange multiplier $\lambda$ so that $\lambda=1$
corresponds to a situation where the relative importance of the first 
and second term in the Lagrangian $\mathcal{L}$ are equal
($w$ is defined as $w=\sum_{\bb}w_{\bb}$ and $w_{\bb}$ are $\kk$-point weights
appearing in the definition of $\Omega$; see Appendix~\ref{sec:appndx-spread}). 

\subsubsection{Replacing \texorpdfstring{$\Omega$}{Omega} with \texorpdfstring{$\OmegaIOD$}{OmegaIOD}}\label{subsec:approx-omega-iod}

Now we show that within our approach one can replace, in
Eq.~\eqref{eq:altsol-opt2}, the
total spread $\Omega$ with $\OmegaIOD(=\OmegaI+\OmegaOD)$ thus ignoring diagonal
part of the spread $\OmegaD$.

We now examine how the diagonal
and off-diagonal spread depend on the gauge transformation written in
the Wannier space. The most general gauge transformation of Bloch states
is given by Eq.~\eqref{eq:gauge} and it involves an arbitrary unitary
transformation of the states at each $\kk$ point in the Brillouin zone.  In
the Wannier space, this same gauge transformation corresponds to the
unitary mixtures of WF's among all unit cells,
\begin{align}
\ket{\mathbf{0} n} \rightarrow 
\sum_{\mathbf{P}}
\sum_m
u^{(\mathbf{P})}_{mn} \ket{\mathbf{P} m}.
\end{align}
Here the matrix $u^{(\mathbf{P})}_{mn}$ is the Fourier transform of the matrix
$u^{(\kk)}_{mn}$ in Eq.~\eqref{eq:gauge}.
A gauge transformation for
which $u^{(\mathbf{P})}_{mn}$ is nonzero only for $\mathbf{P}=\mathbf{0}$ we
will call an intracell gauge transformation, since it involves only
mixtures of the WFs in the same unit cell.

Let us now start from a set of MLWFs in the home cell
$\ket{\mathbf{0} n}$ and see what is the effect of the intracell
gauge transformation on $\OmegaD$ and $\OmegaOD$.  First we will
express the diagonal and off-diagonal spread in terms of the
WFs\cite{PhysRevB.56.12847},
\begin{align}
\OmegaD &= \sum_n \sum_{\mathbf{R}\neq\mathbf{0}}
\qty|\matrixel{\mathbf{R}n}{\rr}{\mathbf{0}n}|^2,
\\
\OmegaOD &= \sum_{m \neq n} \sum_{\mathbf{R}}
\qty|\matrixel{\mathbf{R}m}{\rr}{\mathbf{0}n}|^2.
\end{align}
Since the MLWFs are exponentially localized, we expect that the dominant
term of a gauge dependent spread $\OmegaD+\OmegaOD$ will be the
$\RR=0$ term.  Since the $\RR=0$ term appears only in $\OmegaOD$, it will
dominate over $\OmegaD$ for an intracell gauge
transformation.

Let us return now back to the optimization problem
Eq.~\eqref{eq:altsol-opt2} in question.  By construction, the OPFs
$\bar{g}_j$ approximately overspan the space of MLWFs near the
origin; in other words, they are related by an intracell
gauge transformation.  Therefore, we are justified in ignoring the
diagonal part of the spread $\OmegaD$ in Eq.~\eqref{eq:altsol-opt2}.

With this simplification, the problem of finding MLWFs is reduced to
finding a rectangular semiunitary matrix $W$ and
a real number $\lambda$
which are at a saddle point of the Lagrangian,
\begin{align}
\mathcal{L} \qty(W,\lambda) = &
\OmegaIOD \qty[W^{\dagger}U^{(\kk)\dagger}_{A}m^{(\kk,\bb)}U^{(\kk+\bb)}_{A}W]
\notag \\
+ & \lambda w \sum_{\kk}\sum_{i=1}^N \abs{ \qty[W^{\dagger}A^{(\kk)\dagger}A^{(\kk)}
W]_{ii} - 1 }^2.
\end{align}
Inserting here an explicit definition of $\OmegaIOD$ (see Appendix~\ref{sec:appndx-spread}) and ignoring the constant term and the $1/N_\kk$ prefactor, we obtain
\begin{align}
\mathcal{L} \qty(W,\lambda) = &
-\sum_{\kk,\bb}w_{\bb}\sum_{i=1}^N \abs{
\qty[W^{\dagger}U^{(\kk)\dagger}_{A}m^{(\kk,\bb)}U^{(\kk+\bb)}_{A}W] _{ii}}^{2}
\notag \\
+ & \lambda w \sum_{\kk}\sum_{i=1}^N \abs{ \qty[W^{\dagger}A^{(\kk)\dagger}A^{(\kk)}
W]_{ii} - 1 }^2.
\label{eq:full_lagrangian}
\end{align}
Let us now define the following two quantities that are independent of $W$ and $\lambda$:
\begin{align}
M^{(\kk,\bb)} & = U^{(\kk)\dagger}_{A}m^{(\kk,\bb)}U^{(\kk+\bb)}_{A} ,
\\
S^{(\kk)} & = A^{(\kk)\dagger}A^{(\kk)} - I_M .
\end{align}
With this simplification, the Lagrangian Eq.~\eqref{eq:full_lagrangian} now
simply reads
\begin{align}
\mathcal{L} \qty(W,\lambda) = 
\sum_{\alpha} t^{(\alpha)} \sum_{i=1}^N \abs{
\qty[W^{\dagger}X^{(\alpha)} W] _{ii}}^{2}.
\label{eq:clean_lagrangian}
\end{align}
Here $X^{(\alpha)}$ stands for a collection of $M^{(\kk,\bb)}$ and $S^{(\kk)}$
matrices.
The $t^{(\alpha)}$ are the weights associated with the matrices $X^{(\alpha)}$,
with a weight $-w_{\bb}$ for the $M^{(\kk,\bb)}$ matrices and a weight $\lambda
w$ for the $S^{(\kk)}$ matrices.
Therefore, we have reduced a problem of finding MLWFs to the problem
of codiagonalizing a set of large ($M\!\times\!M$) square matrices
$X^{(\alpha)}$ with a single (i.e. $\kk$-point independent)
rectangular ($M\!\times\!N$) matrix $W$. A mathematically similar
approach for a square matrix $W$ has been used in
Ref.~\onlinecite{Gygi20031} to find MLWFs of a localized system.

In Appendix~\ref{sec:appndx-codiag} we present a numerically efficient 
algorithm for minimizing Eq.~\eqref{eq:clean_lagrangian},
largely following
Refs.~\citenum{doi:10.1137/S0895479893259546} and \citenum{Sørensen2012617}.
In the following section, we illustrate the OPF procedure and empirically
validate the approximations discussed above.

\section{Illustration of our approach}\label{sec:examples}

We now illustrate the OPF procedure on a variety of
systems with chemical bonding ranging from ionic to covalent.
For predominantly ionic materials we choose 
NaCl, Cr$_2$O$_3$, and LaMnO$_3$.  The last two cases have additional complexity
because they have magnetic and orbital order on the transition metals.
For predominantly covalently bonded materials we choose cubic silicon (c-Si), 
strongly distorted silicon with 20 atoms in the primitive unit cell 
(Si-20 from Ref.~\citenum{PhysRevLett.110.118702}), cubic GaAs, and SiO$_2$
in the ideal $\beta$-cristobalite structure.

We computed Bloch wave functions for all seven compounds 
within the density-functional theory
and plane wave pseudopotential approach as implemented 
in the \mbox{{\sc Quantum}} \mbox{{\sc ESPRESSO}} package.\cite{QE-2009} The atomic
potentials were replaced with  ultrasoft\cite{PhysRevB.41.7892} 
pseudopotentials from the GBRV\cite{Garrity2014446} library. 
For the plane wave cutoff, we used 40 and 200~Ry for the wave functions 
and charge density, respectively. All calculations are done with 
experimental lattice parameters. 
In the case of Cr$_2$O$_3$ we sampled the Brillouin zone on 
a uniform $6\times6\times6$ $\kk$-point grid and for all
other cases we used a $4\times4\times4$ grid.

Using the Bloch wave functions, we computed the overlap matrices $m^{(\kk,\bb)}$
between the neighboring Bloch states and the
overlaps $A^{(\kk)}$ between the Bloch states and atomiclike functions that approximately
overspan the space of MLWFs. For predominantly ionic
materials in our test (NaCl, Cr$_2$O$_3$, and LaMnO$_3$), $A^{(\kk)}$ includes
projections of Bloch states into 
all valence atomiclike functions for all atoms in the primitive unit cell.
For covalently bonded materials (c-Si, Si-20, GaAs, and SiO$_2$) 
some Wannier function centers lie on the edge of the primitive 
unit cell (see Sec.~\ref{sec:altsol}), so we included in $A^{(\kk)}$ 
projections onto atoms near the edge of the cell.
Failing to
include these additional projections in the case of c-Si yields
Wannier
functions at the computational unit cell boundary with spreads two times larger
than if we include the additional projections.

We also checked the opposite case by overspanning the space of MLWFs
even further by
including orbitals into $A^{(\kk)}$ that are nominally
not in valence (for example, d-orbitals in the case of cubic silicon).
In this case, the final spread for the WFs for the occupied valence band complex is unaffected and the matrix elements of
$W$ corresponding to these additional orbitals is small, as expected.

Given matrices $m^{(\kk,\bb)}$ and $A^{(\kk)}$ and a choice of the parameter
$\lambda$ we now find matrix $W$ (i.e., OPFs) that minimizes Lagrangian from
Eq.~\eqref{eq:clean_lagrangian} using the algorithm described in 
Appendix~\ref{sec:appndx-codiag}.  Given $W$, we construct the $u^{(\kk)}_{AW}$
to rotate Bloch states into a smooth gauge as described in 
Sec.~\ref{sec:altsol}.  The smoothness of this gauge is quantified
by first computing the spread $\Omega^{\rm OPF}$ from the rotated overlap
matrices in Eq.~\eqref{eq:altsol-opt2} and then comparing it to the
spread $\Omega^{\rm GM}$ at the global minimum.
(We define $\Omega^{\rm GM}$ to be a spread of the Wannier functions after 
running both steps of the standard procedure for obtaining MLWFs.  For convenience,
in the first step of finding the global minimum, we do not guess the initial projections but instead
project into the OPFs obtained from our approach.)

\begin{figure}
  \includegraphics{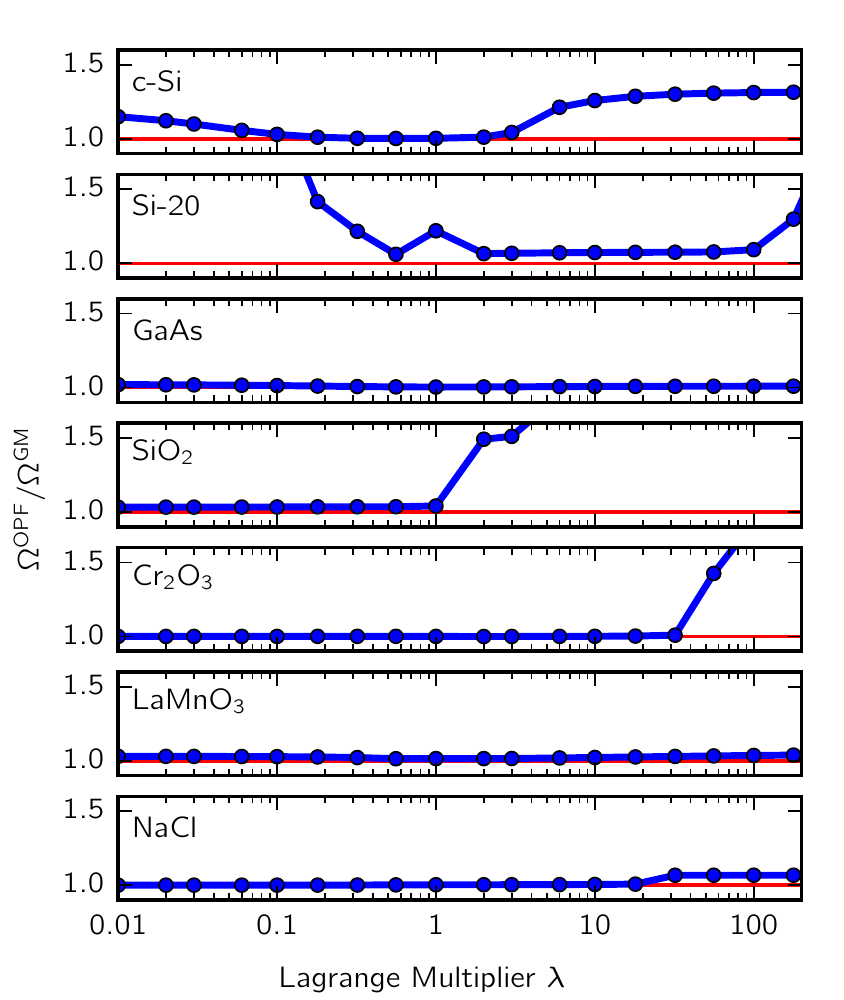}
  \caption{(Color online) Ratio of $\Omega^{\rm OPF}$ and $\Omega^{\rm GM}$ as a function of
  Lagrange multiplier $\lambda$ on a logarithmic scale.
  }
\label{fig:spreads}
\end{figure}

Figure~\ref{fig:spreads} shows, for all seven cases studied, the ratio of the
spread $\Omega^{\rm OPF}$ and $\Omega^{\rm GM}$ as a function of $\lambda$ 
on a logarithmic scale. 
In all cases, the spread $\Omega^{\rm OPF}$
is nearly insensitive to the value of $\lambda$ over several orders of magnitude.
For example, in the case of GaAs or LaMnO$_3$ spread $\Omega^{\rm OPF}$ is nearly the same
for $0.01 < \lambda < 100$.  In the worst case scenario (c-Si), the spread is still
nearly the same for $0.1 < \lambda < 2$.
Therefore, even though in principle one may need to vary $\lambda$ to find
an optimal value of spread, in practice, $\lambda \sim 1$ is usually
a good enough choice.

In each of the seven test cases, the spread 
$\Omega^{\rm OPF}$ is only just 1\% larger than at a global minimum ($\Omega^{\rm GM}$). 
In the worst case situation (Si-20), the spread is only 6\% larger than 
at a global minimum.  As mentioned earlier in Sec.~\ref{sec:altsol}, this spread
could be reduced further by starting from OPFs as initial projections
and running the second step of
the standard procedure.

We give numerical values of $\Omega^{\rm OPF}$ and $\Omega^{\rm GM}$
in Table~\ref{tab:table1} along with a decomposition of spread into
diagonal and off-diagonal components.  From here we find an additional
validation of two simplifications discussed in Sec.~\ref{sec:approx}.  First,
Table~\ref{tab:table1} shows that linearization of $u_{AW}$ is justified
since the off-diagonal component of the spread $\Omega^{\rm OPF}$ and 
$\Omega^{\rm GM}$ is nearly the same.  Second, replacing $\Omega$ with $\OmegaIOD$
(thus, ignoring diagonal spread) is justified within our approach since 
diagonal spread of $\Omega^{\rm OPF}$ and $\Omega^{\rm GM}$ are both very small
compared to the total spread.

\begin{table}
\caption{\label{tab:table1}
  Total spread $\Omega^{\rm OPF}$ computed within our approach and 
  at the global minimum $\Omega^{\rm GM}$ for all seven materials studied.
  We also give diagonal and off-diagonal components of spread in each case ($\OmegaD$ 
  and $\OmegaOD$).
  The spreads $\Omega^{\rm OPF}$ are obtained using the optimal value of
  $\lambda$ (see Fig.~\ref{fig:spreads}). The units for the spreads are \AA$^{2}$.
  In the case of Cr$_{2}$O$_{3}$ we wannierize only the topmost 12 bands below the 
  Fermi level, and in the case of LaMnO$_{3}$ we wannierize the top 2 spin-up bands. 
  In all other cases we wannierize all valence bands.}
\begin{ruledtabular}
\begin{tabular}{lddd.ddd}
\multicolumn{1}{c}{} & \multicolumn{3}{c}{$\Omega^{\mathrm{OPF}}$} && 
\multicolumn{3}{c}{$\Omega^{\mathrm{GM}}$} \\
\cline{2-4} \cline{6-8} &
\multicolumn{1}{c}{\multirow{2}{*}{Total}} & \multicolumn{2}{c}{Components} &&
\multicolumn{1}{c}{\multirow{2}{*}{Total}} & \multicolumn{2}{c}{Components} \\
\cline{3-4} \cline{7-8} & & \multicolumn{1}{c}{$\OmegaD$} & \multicolumn{1}{c}{$\OmegaOD$} && & \multicolumn{1}{c}{$\OmegaD$} & \multicolumn{1}{c}{$\OmegaOD$} \\
\hline
c-Si            & 6.51   & 0.00 & 0.59   && 6.48  & 0.00 & 0.56  \\
Si-20           & 103.91 & 0.05 & 14.85  && 97.59 & 0.04 & 8.54  \\
GaAs            & 7.25   & 0.02 & 0.61   && 7.22  & 0.01 & 0.59  \\
SiO$_{2}$       & 9.39   & 0.00 & 1.98   && 9.18  & 0.00 & 1.78  \\
Cr$_{2}$O$_{3}$ & 36.04  & 0.10 & 1.17   && 35.74 & 0.05 & 0.91  \\
LaMnO$_{3}$     & 14.89  & 0.15 & 0.17   && 14.68 & 0.00 & 0.11  \\
NaCl            & 4.05   & 0.00 & 0.80   && 4.04  & 0.00 & 0.79  \\
\end{tabular}
\end{ruledtabular}
\end{table}

\subsection{Insight gained from the matrix \texorpdfstring{$W$}{W}}

To demonstrate the kind of insight that can be gained from analyzing the $W$ matrix, we
analyze here in more detail case of LaMnO$_3$ and Cr$_2$O$_3$. In both cases, $s$ and $p$ orbitals on
the neighboring oxygen atoms outside the computational
unit cell are included in $A^{(\kk)}$ in order to complete
the octahedral coordination of the Cr and Mn atoms.

\begin{figure}
  \includegraphics[width=3.4in]{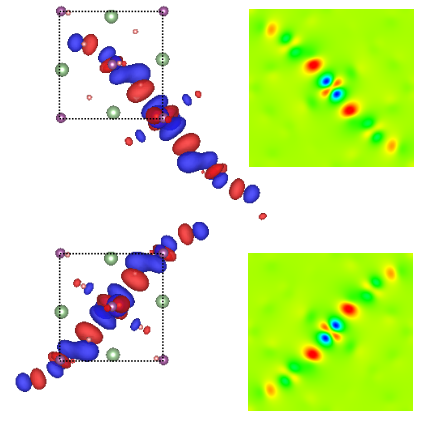}
  \caption{(Color online) Maximally localized Wannier functions of the two topmost valence
  bands in LaMnO$_3$.  Isosurfaces of the WFs are shown on the left, looking down along the $c$ axis.
  The large green dots are La,
  medium purple dots are Mn, and small red dots are O.  On the right, we show contour plots
  of the Wannier functions in the plane perpendicular to the $c$ axis, cutting through the Mn atom.}
\label{fig:lamno3}
\end{figure}

We studied LaMnO$_3$ in its low temperature
$(\lesssim 135\textrm{K})$ A-AFM phase characterized by ferromagnetic
ordering of the Mn spins in-plane and antiferromagnetic order
between planes.\cite{Elemans1971238} In addition to the magnetic order, Mn
$d$ states are orbitally ordered, oxygen octahedra are tilted and Jahn-Teller
distorted.
In the following, we focus only on the two topmost spin-polarized bands 
below the Fermi level.
Analyzing the $W$ matrix
we see that the Wannier functions for the two topmost bands in LaMnO$_3$
are dominantly composed of rotated $d_{z^{2}}$ components on Mn that are 
oriented perpendicular to each other.
This can be seen also by analyzing the $W$ matrix for these two WFs,
\begin{align*}
  \ket{1} &\approx 0.5\ket{\mathrm{Mn1};d_{z^{2}}} + 0.6\ket{\mathrm{Mn1};d_{xy}} \\
  \ket{2} &\approx 0.6\ket{\mathrm{Mn2};d_{z^{2}}} - 0.5\ket{\mathrm{Mn2};d_{xy}}.
\end{align*}
Figure \ref{fig:lamno3} shows a plot of these WFs for the top bands 
with isosurfaces
in the left panels and contour plots in the right panels. 
The
contours are plotted in the plane perpendicular to the $c$ axis, cutting through
the Mn atom.

Furthermore, The $W$ matrix shows hybridization of the Mn
$d$ states with the oxygen $p$ states, with the corresponding elements of $W$
having a magnitude of approximately 0.2 (three times smaller than for Mn $d$ states).
The contribution of the
$p$-like lobes (colored red) can be seen in the right panels of
Fig.~\ref{fig:lamno3} as the large lobes near the center.

Now we analyze the case of Cr$_2$O$_3$ in more detail. Cr$_2$O$_3$ is an antiferromagnetic
insulator with four Cr atoms in the primitive unit cell. Therefore, we
expect each Cr$^{3+}$ ion to nominally have three occupied $d$ orbitals of same spin.
These three occupied $d$ orbitals on four Cr ions 
form a complex of $3\times4=12$ isolated bands that
make up the topmost valence bands.
Again, analyzing the $W$ matrix we obtained within our approach we find that each
of the twelve WFs is a particular linear combination of all five $d$ orbitals,
all having the same spin component along the $z$ axis.
In fact, there is a large degeneracy regarding the particular combination of
$d$ orbitals that make up the WFs.  For example, even slight change in $\lambda$
from 1 to 2
gives different linear combinations of $d$ orbitals, while the spread remains
nearly the same (see Fig.~\ref{fig:spreads}).
This observation is consistent with the fact that the choice of MLWFs is not always
unique.  This was first suggested in Ref.~\onlinecite{PhysRevB.56.12847} 
for the case of LiCl. There it was found that an arbitrary rotation of the $sp^3$ 
orbitals on chlorine atoms has no effect on the total spread $\Omega$.

\section{Summary}\label{sec:summary}

We present an automated procedure for constructing maximally localized Wannier
functions for an isolated group of bands.
The extension of our method to the case of entangled bands will be the subject
of future work.

Instead of having to guess functions 
(initial projections) that approximate the MLWFs as in Ref.~\onlinecite{PhysRevB.56.12847}, our 
approach only requires as input a set of functions that overspan the space
of MLWFs.  In practice, this can rather easily be achieved by selecting
an appropriate set of valence atomiclike functions.

\begin{acknowledgments}
We thank David Vanderbilt for discussion.
This research was supported by the Theory Program at the Lawrence Berkeley
National Lab through the Office of Basic Energy Sciences, U.S. Department of
Energy under Contract No. DE-AC02-05CH11231 (methods and algorithm
developments), and by the National Science Foundation under Grant No.
DMR15-1508412 (band structure calculations). Computational resources have been
provided by the National Energy Research Scientific Computing Center, which is
supported by the Office of Science of the U.S. Department of Energy.
\end{acknowledgments}

\appendix

\section{Spread functional}\label{sec:appndx-spread}

Here we express components of the spread $\Omega$, corresponding to $N$ 
composite bands, as a function of the
overlap matrices $m^{(\kk,\bb)}$ following
Ref.~\onlinecite{PhysRevB.56.12847},
\begin{align}\label{eq:spread-all}
  \begin{split}
  \OmegaI &= \frac{1}{N_{\kk}}\sum_{\kk,\bb}w_{\bb}\qty(N-\sum_{ij}\abs{m^{(\kk,\bb)}_{ij}}^{2}), \\
  \OmegaD &= \frac{1}{N_{\kk}}\sum_{\kk,\bb}w_{\bb}\sum_{i}\qty(-\textrm{Im}\,\ln m^{(\kk,\bb)}_{ii}-\bb\cdot\overline{\rr}_i)^2, \\
  \OmegaOD &= \frac{1}{N_{\kk}}\sum_{\kk,\bb}w_{\bb}\sum_{i\neq j}\abs{m^{(\kk,\bb)}_{ij}}^{2}.
  \end{split}
\end{align}
The $w_{\bb}$ are the weights of the $\bb$ vectors connecting neighboring
$\kk$ points (see Sec.~2.1 of Ref.~\citenum{Mostofi2008685}), while
\begin{equation}
  \overline{\rr}_i = \frac{1}{N_{\kk}}\sum_{\kk,\bb}w_{\bb}\bb\Im\,\ln m^{(\kk,\bb)}_{ii}.
\end{equation}

We note that the diagonal and off-diagonal parts of the spread depend
only on the diagonal and off-diagonal components of the overlap
matrices, respectively.  Combining the invariant and off-diagonal
parts of the spread gives an expression that depends only on the
diagonal components of the overlap matrices,
\begin{align}\label{eq:spread-iod}
  \begin{split}
    \OmegaIOD &= \OmegaI+\OmegaOD \\
              &= \frac{1}{N_\kk}\sum_{\kk,\bb}w_{\bb}\sum_{i=1}^N\qty[1-\abs{m^{(\kk,\bb)}_{ii}}^{2}].
  \end{split}
\end{align}

\section{Codiagonalization algorithm}\label{sec:appndx-codiag}

In the main text, the construction of localized Wannier functions is recast
into the following mathematical problem.  Given a set of large 
$M\!\times\!M$ matrices $X^{(\alpha)}$, we wish to  find
a single rectangular semiunitary $M\!\times\!N$ matrix $W$ such that 
the set of small $N\!\times\!N$ matrices $W^{\dagger} X^{(\alpha)} W$
minimize the Lagrangian $\mathcal{L}$, defined in 
Eq.~\eqref{eq:clean_lagrangian},
\begin{align}
\sum_{\alpha} t^{(\alpha)} \sum_{i=1}^N \abs{
\qty[W^{\dagger}X^{(\alpha)} W] _{ii}}^{2}.
\end{align}
We parametrize the semiunitary matrix $W$ as follows.
First, we define $W$ to be first $N$ columns of an $M\!\times\!M$ unitary matrix
$\widetilde{W}$.  Second, we iteratively parametrize the enlarged matrix
$\widetilde{W}$ as a product (post-multiplication) of Givens
rotations,\cite{Golub:1996:MC:248979}
\begin{equation}
  \widetilde{W} = \prod_{l=1}^{L}\prod_{i=1}^{N}\prod_{j=i+1}^{M} R_{l}[i,j,\theta,\phi].
\label{eq:prod_givens}
\end{equation}
Here integer $l$ denotes a particular iteration in the expansion. 

A Givens rotation $R[i,j,\theta,\phi]$
is the most general unitary matrix that acts only on $i$-th and $j$-th rows
and columns.  Therefore, we parametrize $R[i,j,\theta,\phi]$ with two
angles $\theta$ and $\phi$ 
as a matrix equal to the
identity matrix for all elements except for the $ii$, $ij$, $ji$, and $jj$ elements,
\begin{equation}
\begin{pmatrix}
R_{ii} & R_{ij} \\
R_{ji} & R_{jj} 
\end{pmatrix}
=
\begin{pmatrix}
\cos\theta & e^{i\phi} \sin\theta \\
-e^{-i\phi} \sin\theta & \cos\theta 
\end{pmatrix}.
\end{equation}
The only diagonal elements of $X^{(\alpha)}$ affected by
$R[i,j,\theta,\phi]$ are $X^{(\alpha)}_{ii}$ and $X^{(\alpha)}_{jj}$.  
Therefore there is no need to include in
Eq.~\eqref{eq:prod_givens} cases when both $i$ and $j$ are larger
than $N$, since that operation will have no effect on the Lagrangian.
In addition, we don't consider cases when $j<i$ since that
transformation is captured by $j>i$.
With this parametrization an arbitrary unitary matrix $\widetilde{W}$ can 
be approximated to an arbitrary precision
with large enough number of iterations, $L$.

Let us now see how does a single Givens rotation affect the Lagrangian.
For a Givens rotation $R[i,j,\theta,\phi]$, the sum of the weighted square moduli of the
diagonal elements ($ii$ and $jj$) of a set of rotated matrices
$R^{\dagger}X^{(\alpha)}R$ are\cite{Sørensen2012617}
\begin{align}
  \sum_{\alpha} t^{(\alpha)} \qty|\qty[R^{\dagger}X^{(\alpha)}R]_{ii}|^{2} &= \mathbf{x}^{\intercal}Q\mathbf{x} + \mathbf{p}^{\intercal}\mathbf{x} + c \label{eq:Xii} \\
  \sum_{\alpha} t^{(\alpha)} \qty|\qty[R^{\dagger}X^{(\alpha)}R]_{jj}|^{2} &= \mathbf{x}^{\intercal}Q\mathbf{x} - \mathbf{p}^{\intercal}\mathbf{x} + c \label{eq:Xjj}
\end{align}
where
\begin{equation}\label{eq:appndx-x}
  \mathbf{x}^{\intercal}=\qty(\cos 2\theta,\sin 2\theta\cos\phi,\sin 2\theta\sin\phi)
\end{equation}
is a vector with unit norm by construction. 
The coefficients of the quadratic forms above [Eqs.~\eqref{eq:Xii} and \eqref{eq:Xjj}] depend only
on the $ii$, $ij$, $ji$, and $jj$ components of the $X^{(\alpha)}$ matrices
\begin{align}
  \begin{split}
    Q          &= \sum_{\alpha} t^{(\alpha)} \Re\qty[\mathbf{z}^{(\alpha)}\mathbf{z}^{(\alpha)\dagger}] \\
    \mathbf{p} &= \sum_{\alpha} t^{(\alpha)} \Re\qty[\qty(X^{(\alpha)}_{ii}+X^{(\alpha)}_{jj})^{*}\mathbf{z}^{(\alpha)}] \\
    c          &= \sum_{\alpha} \frac{1}{4} t^{(\alpha)} \qty|X^{(\alpha)}_{ii}+X^{(\alpha)}_{jj}|^2
  \end{split}
\end{align}
where
\begin{equation}
  \mathbf{z}^{(\alpha)} = \frac{1}{2}
  \begin{bmatrix}
    X^{(\alpha)}_{ii} - X^{(\alpha)}_{jj} \\
    -\qty(X^{(\alpha)}_{ij} + X^{(\alpha)}_{ji}) \\
    i\qty(X^{(\alpha)}_{ij} - X^{(\alpha)}_{ji}) \\
  \end{bmatrix}.
\end{equation}

We now consider two cases. First, if $j \leq N$ both the $ii$ and $jj$ diagonal
elements enter the Lagrangian $\mathcal{L}$ so we need to find $\mathbf{x}$
that minimizes the sum of Eqs.~\eqref{eq:Xii} and \eqref{eq:Xjj}, 
\begin{equation}
\begin{split}
  \sum_{\alpha} & \, t^{(\alpha)}\qty|\qty[R^{\dagger}X^{(\alpha)}R]_{ii}|^{2} + t^{(\alpha)}\qty|\qty[R^{\dagger}X^{(\alpha)}R]_{jj}|^{2} \\
  & = 2\mathbf{x}^{\intercal}Q\mathbf{x} + 2c.
\end{split}
\end{equation}
This is a quadratic programming problem with the constraint that
$\abs{\mathbf{x}}=1$.  Here, the Lagrangian is simply minimized for
$\mathbf{x}$ that is the normalized eigenvector corresponding to the
minimal eigenvalue of $Q$. For numerical stability, if the first
component of $\mathbf{x}$ (i.e., $\cos 2\theta$) happens to be
negative
we choose $-\mathbf{x}$ instead of $\mathbf{x}$.

In the second case $(j\!>\!N)$, only the $ii$ diagonal components 
enters the Lagrangian $\mathcal{L}$
so we need to find $\mathbf{x}$ that minimizes Eq.~\eqref{eq:Xii},
\begin{equation}\label{eq:case2}
  \mathbf{x}^{\intercal}Q\mathbf{x} + \mathbf{p}^{\intercal}\mathbf{x} + c.  \\
\end{equation}
The solution of this problem is discussed in Ref.~\citenum{Sørensen2012617}
within the context of matrix codiagonalization.  However we find the general
quadratic programming solution from Ref.~\citenum{Gander1989815}
more numerically stable. 
Following Ref.~\citenum{Gander1989815},
we first find the minimal eigenvalue $\chi_{\mathrm{min}}$
of the quadratic eigenvalue problem (QEP)
\begin{equation}\label{eq:qep}
  \qty(\chi^2 A_2 + \chi A_1 + A_0)\mathbf{x} = 0,
\end{equation}
with
\begin{align}
  \begin{split}
  & A_2 = I_3 \\
  & A_1 = -2Q \\
  & A_0 = Q^2 - \frac{1}{4}\mathbf{p}\mathbf{p}^{\intercal}.
  \end{split}
\end{align}
The QEP is linearized by introducing
\begin{equation}
  \widetilde{\mathbf{x}} = \begin{pmatrix}\chi \mathbf{x}\\\mathbf{x}\end{pmatrix}
\end{equation}
yielding a generalized eigenvalue problem
\begin{equation}
  A\widetilde{\mathbf{x}} = \chi B\widetilde{\mathbf{x}},
\end{equation}
with
\begin{align}
  \begin{split}
  & A = \mqty*(A_1 && A_0 \\ -I_3 && 0) \\
  & B = \mqty*(A_2 && 0 \\ 0 && I_3).
  \end{split}
\end{align}
This generalized eigenvalue problem we solve 
using standard linear algebra techniques.
The solution $\mathbf{x}$ that minimizes Eq.~\ref{eq:case2} depends on whether
$\chi_{\mathrm{min}}$ is in the spectrum of $Q$ or not.

If $\chi_{\mathrm{min}}$ is not in the spectrum (i.e. not an
eigenvalue) of $Q$ then the solution is
$\mathbf{x}=\qty(Q-\chi_{\mathrm{min}}I)^{-1}(-\mathbf{p}/2)$.
If $\chi_{\mathrm{min}}$ is an eigenvalue of $Q$ we first define
\begin{equation}
\mathbf{u}:=\qty(Q-\chi_{\mathrm{min}}I)^{+}\qty(-\mathbf{p}/2).
\end{equation}
Here symbol $^{+}$ denotes a matrix pseudoinverse.
A nontrivial solution to Eq.~\ref{eq:case2} exists only when the following conditions are satisfied:
\begin{equation}
  (Q-\chi_{\mathrm{min}}I)\mathbf{u}=-\mathbf{p}/2 \quad \mathrm{and} \quad \abs{\mathbf{u}}\leq1.
\end{equation}
Finally, if $\abs{\mathbf{u}}=1$, then the solution is
$\mathbf{x}=\mathbf{u}$. Otherwise
($\abs{\mathbf{u}}<1$) the solution is
$\mathbf{x}=\mathbf{u}+\bm{\xi}$. Here $\bm{\xi}$ is an
eigenvector of $Q$ corresponding to $\chi_{\mathrm{min}}$ 
chosen so that 
$\abs{\bm{\xi}}^2=1-\abs{\mathbf{u}}^2$.

Once the $\mathbf{x}$ is found for a given $(i,j)$ in
either of the two approaches we determine the
corresponding angles $(\theta,\phi)$ from Eq.~\eqref{eq:appndx-x} 
and construct the Givens rotation $R[i,j,\theta,\phi]$. 
Next we update at each iteration the matrix $\widetilde{W}$
according to the postmultiplication 
parametrization from Eq.~\eqref{eq:prod_givens},
\begin{equation}
  \begin{split}
  \widetilde{W} &\rightarrow \widetilde{W}R.
  \end{split}
\end{equation}
This iterative procedure over $i$, $j$, and $l$ 
continues until the Lagrangian converges.

\bibliography{refs}

\begin{thebibliography}{23}%
\makeatletter
\providecommand \@ifxundefined [1]{%
 \@ifx{#1\undefined}
}%
\providecommand \@ifnum [1]{%
 \ifnum #1\expandafter \@firstoftwo
 \else \expandafter \@secondoftwo
 \fi
}%
\providecommand \@ifx [1]{%
 \ifx #1\expandafter \@firstoftwo
 \else \expandafter \@secondoftwo
 \fi
}%
\providecommand \natexlab [1]{#1}%
\providecommand \enquote  [1]{``#1''}%
\providecommand \bibnamefont  [1]{#1}%
\providecommand \bibfnamefont [1]{#1}%
\providecommand \citenamefont [1]{#1}%
\providecommand \href@noop [0]{\@secondoftwo}%
\providecommand \href [0]{\begingroup \@sanitize@url \@href}%
\providecommand \@href[1]{\@@startlink{#1}\@@href}%
\providecommand \@@href[1]{\endgroup#1\@@endlink}%
\providecommand \@sanitize@url [0]{\catcode `\\12\catcode `\$12\catcode
  `\&12\catcode `\#12\catcode `\^12\catcode `\_12\catcode `\%12\relax}%
\providecommand \@@startlink[1]{}%
\providecommand \@@endlink[0]{}%
\providecommand \url  [0]{\begingroup\@sanitize@url \@url }%
\providecommand \@url [1]{\endgroup\@href {#1}{\urlprefix }}%
\providecommand \urlprefix  [0]{URL }%
\providecommand \Eprint [0]{\href }%
\providecommand \doibase [0]{http://dx.doi.org/}%
\providecommand \selectlanguage [0]{\@gobble}%
\providecommand \bibinfo  [0]{\@secondoftwo}%
\providecommand \bibfield  [0]{\@secondoftwo}%
\providecommand \translation [1]{[#1]}%
\providecommand \BibitemOpen [0]{}%
\providecommand \bibitemStop [0]{}%
\providecommand \bibitemNoStop [0]{.\EOS\space}%
\providecommand \EOS [0]{\spacefactor3000\relax}%
\providecommand \BibitemShut  [1]{\csname bibitem#1\endcsname}%
\let\auto@bib@innerbib\@empty
\bibitem [{\citenamefont {Marzari}\ \emph {et~al.}(2012)\citenamefont
  {Marzari}, \citenamefont {Mostofi}, \citenamefont {Yates}, \citenamefont
  {Souza},\ and\ \citenamefont {Vanderbilt}}]{RevModPhys.84.1419}%
  \BibitemOpen
  \bibfield  {author} {\bibinfo {author} {\bibfnamefont {N.}~\bibnamefont
  {Marzari}}, \bibinfo {author} {\bibfnamefont {A.~A.}\ \bibnamefont
  {Mostofi}}, \bibinfo {author} {\bibfnamefont {J.~R.}\ \bibnamefont {Yates}},
  \bibinfo {author} {\bibfnamefont {I.}~\bibnamefont {Souza}}, \ and\ \bibinfo
  {author} {\bibfnamefont {D.}~\bibnamefont {Vanderbilt}},\ }\href {\doibase
  10.1103/RevModPhys.84.1419} {\bibfield  {journal} {\bibinfo  {journal} {Rev.
  Mod. Phys.}\ }\textbf {\bibinfo {volume} {84}},\ \bibinfo {pages} {1419}
  (\bibinfo {year} {2012})}\BibitemShut {NoStop}%
\bibitem [{\citenamefont {Wu}\ \emph {et~al.}(2006)\citenamefont {Wu},
  \citenamefont {Di\'eguez}, \citenamefont {Rabe},\ and\ \citenamefont
  {Vanderbilt}}]{PhysRevLett.97.107602}%
  \BibitemOpen
  \bibfield  {author} {\bibinfo {author} {\bibfnamefont {X.}~\bibnamefont
  {Wu}}, \bibinfo {author} {\bibfnamefont {O.}~\bibnamefont {Di\'eguez}},
  \bibinfo {author} {\bibfnamefont {K.~M.}\ \bibnamefont {Rabe}}, \ and\
  \bibinfo {author} {\bibfnamefont {D.}~\bibnamefont {Vanderbilt}},\ }\href
  {\doibase 10.1103/PhysRevLett.97.107602} {\bibfield  {journal} {\bibinfo
  {journal} {Phys. Rev. Lett.}\ }\textbf {\bibinfo {volume} {97}},\ \bibinfo
  {pages} {107602} (\bibinfo {year} {2006})}\BibitemShut {NoStop}%
\bibitem [{\citenamefont {Yates}\ \emph {et~al.}(2007)\citenamefont {Yates},
  \citenamefont {Wang}, \citenamefont {Vanderbilt},\ and\ \citenamefont
  {Souza}}]{PhysRevB.75.195121}%
  \BibitemOpen
  \bibfield  {author} {\bibinfo {author} {\bibfnamefont {J.~R.}\ \bibnamefont
  {Yates}}, \bibinfo {author} {\bibfnamefont {X.}~\bibnamefont {Wang}},
  \bibinfo {author} {\bibfnamefont {D.}~\bibnamefont {Vanderbilt}}, \ and\
  \bibinfo {author} {\bibfnamefont {I.}~\bibnamefont {Souza}},\ }\href
  {\doibase 10.1103/PhysRevB.75.195121} {\bibfield  {journal} {\bibinfo
  {journal} {Phys. Rev. B}\ }\textbf {\bibinfo {volume} {75}},\ \bibinfo
  {pages} {195121} (\bibinfo {year} {2007})}\BibitemShut {NoStop}%
\bibitem [{\citenamefont {Giustino}\ \emph {et~al.}(2007)\citenamefont
  {Giustino}, \citenamefont {Cohen},\ and\ \citenamefont
  {Louie}}]{PhysRevB.76.165108}%
  \BibitemOpen
  \bibfield  {author} {\bibinfo {author} {\bibfnamefont {F.}~\bibnamefont
  {Giustino}}, \bibinfo {author} {\bibfnamefont {M.~L.}\ \bibnamefont {Cohen}},
  \ and\ \bibinfo {author} {\bibfnamefont {S.~G.}\ \bibnamefont {Louie}},\
  }\href {\doibase 10.1103/PhysRevB.76.165108} {\bibfield  {journal} {\bibinfo
  {journal} {Phys. Rev. B}\ }\textbf {\bibinfo {volume} {76}},\ \bibinfo
  {pages} {165108} (\bibinfo {year} {2007})}\BibitemShut {NoStop}%
\bibitem [{\citenamefont {Noffsinger}\ \emph {et~al.}(2010)\citenamefont
  {Noffsinger}, \citenamefont {Giustino}, \citenamefont {Malone}, \citenamefont
  {Park}, \citenamefont {Louie},\ and\ \citenamefont
  {Cohen}}]{Noffsinger20102140}%
  \BibitemOpen
  \bibfield  {author} {\bibinfo {author} {\bibfnamefont {J.}~\bibnamefont
  {Noffsinger}}, \bibinfo {author} {\bibfnamefont {F.}~\bibnamefont
  {Giustino}}, \bibinfo {author} {\bibfnamefont {B.~D.}\ \bibnamefont
  {Malone}}, \bibinfo {author} {\bibfnamefont {C.-H.}\ \bibnamefont {Park}},
  \bibinfo {author} {\bibfnamefont {S.~G.}\ \bibnamefont {Louie}}, \ and\
  \bibinfo {author} {\bibfnamefont {M.~L.}\ \bibnamefont {Cohen}},\ }\href
  {\doibase 10.1016/j.cpc.2010.08.027} {\bibfield  {journal} {\bibinfo
  {journal} {Comput. Phys. Commun.}\ }\textbf {\bibinfo {volume} {181}},\
  \bibinfo {pages} {2140 } (\bibinfo {year} {2010})}\BibitemShut {NoStop}%
\bibitem [{\citenamefont {Pizzi}\ \emph {et~al.}(2014)\citenamefont {Pizzi},
  \citenamefont {Volja}, \citenamefont {Kozinsky}, \citenamefont {Fornari},\
  and\ \citenamefont {Marzari}}]{Pizzi2014422}%
  \BibitemOpen
  \bibfield  {author} {\bibinfo {author} {\bibfnamefont {G.}~\bibnamefont
  {Pizzi}}, \bibinfo {author} {\bibfnamefont {D.}~\bibnamefont {Volja}},
  \bibinfo {author} {\bibfnamefont {B.}~\bibnamefont {Kozinsky}}, \bibinfo
  {author} {\bibfnamefont {M.}~\bibnamefont {Fornari}}, \ and\ \bibinfo
  {author} {\bibfnamefont {N.}~\bibnamefont {Marzari}},\ }\href {\doibase
  10.1016/j.cpc.2013.09.015} {\bibfield  {journal} {\bibinfo  {journal}
  {Comput. Phys. Commun.}\ }\textbf {\bibinfo {volume} {185}},\ \bibinfo
  {pages} {422 } (\bibinfo {year} {2014})}\BibitemShut {NoStop}%
\bibitem [{\citenamefont {Marzari}\ and\ \citenamefont
  {Vanderbilt}(1997)}]{PhysRevB.56.12847}%
  \BibitemOpen
  \bibfield  {author} {\bibinfo {author} {\bibfnamefont {N.}~\bibnamefont
  {Marzari}}\ and\ \bibinfo {author} {\bibfnamefont {D.}~\bibnamefont
  {Vanderbilt}},\ }\href {\doibase 10.1103/PhysRevB.56.12847} {\bibfield
  {journal} {\bibinfo  {journal} {Phys. Rev. B}\ }\textbf {\bibinfo {volume}
  {56}},\ \bibinfo {pages} {12847} (\bibinfo {year} {1997})}\BibitemShut
  {NoStop}%
\bibitem [{\citenamefont {Souza}\ \emph {et~al.}(2001)\citenamefont {Souza},
  \citenamefont {Marzari},\ and\ \citenamefont
  {Vanderbilt}}]{PhysRevB.65.035109}%
  \BibitemOpen
  \bibfield  {author} {\bibinfo {author} {\bibfnamefont {I.}~\bibnamefont
  {Souza}}, \bibinfo {author} {\bibfnamefont {N.}~\bibnamefont {Marzari}}, \
  and\ \bibinfo {author} {\bibfnamefont {D.}~\bibnamefont {Vanderbilt}},\
  }\href {\doibase 10.1103/PhysRevB.65.035109} {\bibfield  {journal} {\bibinfo
  {journal} {Phys. Rev. B}\ }\textbf {\bibinfo {volume} {65}},\ \bibinfo
  {pages} {035109} (\bibinfo {year} {2001})}\BibitemShut {NoStop}%
\bibitem [{\citenamefont {Mostofi}\ \emph {et~al.}(2008)\citenamefont
  {Mostofi}, \citenamefont {Yates}, \citenamefont {Lee}, \citenamefont {Souza},
  \citenamefont {Vanderbilt},\ and\ \citenamefont {Marzari}}]{Mostofi2008685}%
  \BibitemOpen
  \bibfield  {author} {\bibinfo {author} {\bibfnamefont {A.~A.}\ \bibnamefont
  {Mostofi}}, \bibinfo {author} {\bibfnamefont {J.~R.}\ \bibnamefont {Yates}},
  \bibinfo {author} {\bibfnamefont {Y.-S.}\ \bibnamefont {Lee}}, \bibinfo
  {author} {\bibfnamefont {I.}~\bibnamefont {Souza}}, \bibinfo {author}
  {\bibfnamefont {D.}~\bibnamefont {Vanderbilt}}, \ and\ \bibinfo {author}
  {\bibfnamefont {N.}~\bibnamefont {Marzari}},\ }\href {\doibase
  10.1016/j.cpc.2007.11.016} {\bibfield  {journal} {\bibinfo  {journal}
  {Comput. Phys. Commun.}\ }\textbf {\bibinfo {volume} {178}},\ \bibinfo
  {pages} {685 } (\bibinfo {year} {2008})}\BibitemShut {NoStop}%
\bibitem [{\citenamefont {Jain}\ \emph {et~al.}(2013)\citenamefont {Jain},
  \citenamefont {Ong}, \citenamefont {Hautier}, \citenamefont {Chen},
  \citenamefont {Richards}, \citenamefont {Dacek}, \citenamefont {Cholia},
  \citenamefont {Gunter}, \citenamefont {Skinner}, \citenamefont {Ceder},\ and\
  \citenamefont {Persson}}]{Jain2013}%
  \BibitemOpen
  \bibfield  {author} {\bibinfo {author} {\bibfnamefont {A.}~\bibnamefont
  {Jain}}, \bibinfo {author} {\bibfnamefont {S.~P.}\ \bibnamefont {Ong}},
  \bibinfo {author} {\bibfnamefont {G.}~\bibnamefont {Hautier}}, \bibinfo
  {author} {\bibfnamefont {W.}~\bibnamefont {Chen}}, \bibinfo {author}
  {\bibfnamefont {W.~D.}\ \bibnamefont {Richards}}, \bibinfo {author}
  {\bibfnamefont {S.}~\bibnamefont {Dacek}}, \bibinfo {author} {\bibfnamefont
  {S.}~\bibnamefont {Cholia}}, \bibinfo {author} {\bibfnamefont
  {D.}~\bibnamefont {Gunter}}, \bibinfo {author} {\bibfnamefont
  {D.}~\bibnamefont {Skinner}}, \bibinfo {author} {\bibfnamefont
  {G.}~\bibnamefont {Ceder}}, \ and\ \bibinfo {author} {\bibfnamefont {K.~A.}\
  \bibnamefont {Persson}},\ }\href {\doibase 10.1063/1.4812323} {\bibfield
  {journal} {\bibinfo  {journal} {APL Mater.}\ }\textbf {\bibinfo {volume}
  {1}},\ \bibinfo {pages} {011002} (\bibinfo {year} {2013})}\BibitemShut
  {NoStop}%
\bibitem [{\citenamefont {Thygesen}\ \emph
  {et~al.}(2005{\natexlab{a}})\citenamefont {Thygesen}, \citenamefont
  {Hansen},\ and\ \citenamefont {Jacobsen}}]{PhysRevLett.94.026405}%
  \BibitemOpen
  \bibfield  {author} {\bibinfo {author} {\bibfnamefont {K.~S.}\ \bibnamefont
  {Thygesen}}, \bibinfo {author} {\bibfnamefont {L.~B.}\ \bibnamefont
  {Hansen}}, \ and\ \bibinfo {author} {\bibfnamefont {K.~W.}\ \bibnamefont
  {Jacobsen}},\ }\href {\doibase 10.1103/PhysRevLett.94.026405} {\bibfield
  {journal} {\bibinfo  {journal} {Phys. Rev. Lett.}\ }\textbf {\bibinfo
  {volume} {94}},\ \bibinfo {pages} {026405} (\bibinfo {year}
  {2005}{\natexlab{a}})}\BibitemShut {NoStop}%
\bibitem [{\citenamefont {Thygesen}\ \emph
  {et~al.}(2005{\natexlab{b}})\citenamefont {Thygesen}, \citenamefont
  {Hansen},\ and\ \citenamefont {Jacobsen}}]{PhysRevB.72.125119}%
  \BibitemOpen
  \bibfield  {author} {\bibinfo {author} {\bibfnamefont {K.~S.}\ \bibnamefont
  {Thygesen}}, \bibinfo {author} {\bibfnamefont {L.~B.}\ \bibnamefont
  {Hansen}}, \ and\ \bibinfo {author} {\bibfnamefont {K.~W.}\ \bibnamefont
  {Jacobsen}},\ }\href {\doibase 10.1103/PhysRevB.72.125119} {\bibfield
  {journal} {\bibinfo  {journal} {Phys. Rev. B}\ }\textbf {\bibinfo {volume}
  {72}},\ \bibinfo {pages} {125119} (\bibinfo {year}
  {2005}{\natexlab{b}})}\BibitemShut {NoStop}%
\bibitem [{\citenamefont {Sakuma}(2013)}]{PhysRevB.87.235109}%
  \BibitemOpen
  \bibfield  {author} {\bibinfo {author} {\bibfnamefont {R.}~\bibnamefont
  {Sakuma}},\ }\href {\doibase 10.1103/PhysRevB.87.235109} {\bibfield
  {journal} {\bibinfo  {journal} {Phys. Rev. B}\ }\textbf {\bibinfo {volume}
  {87}},\ \bibinfo {pages} {235109} (\bibinfo {year} {2013})}\BibitemShut
  {NoStop}%
\bibitem [{\citenamefont {Gygi}\ \emph {et~al.}(2003)\citenamefont {Gygi},
  \citenamefont {Fattebert},\ and\ \citenamefont {Schwegler}}]{Gygi20031}%
  \BibitemOpen
  \bibfield  {author} {\bibinfo {author} {\bibfnamefont {F.}~\bibnamefont
  {Gygi}}, \bibinfo {author} {\bibfnamefont {J.-L.}\ \bibnamefont {Fattebert}},
  \ and\ \bibinfo {author} {\bibfnamefont {E.}~\bibnamefont {Schwegler}},\
  }\href {\doibase 10.1016/S0010-4655(03)00315-1} {\bibfield  {journal}
  {\bibinfo  {journal} {Comput. Phys. Commun.}\ }\textbf {\bibinfo {volume}
  {155}},\ \bibinfo {pages} {1 } (\bibinfo {year} {2003})}\BibitemShut
  {NoStop}%
\bibitem [{\citenamefont {Cardoso}\ and\ \citenamefont
  {Souloumiac}(1996)}]{doi:10.1137/S0895479893259546}%
  \BibitemOpen
  \bibfield  {author} {\bibinfo {author} {\bibfnamefont {J.-F.}\ \bibnamefont
  {Cardoso}}\ and\ \bibinfo {author} {\bibfnamefont {A.}~\bibnamefont
  {Souloumiac}},\ }\href {\doibase 10.1137/S0895479893259546} {\bibfield
  {journal} {\bibinfo  {journal} {SIAM J. Matrix Anal. and Appl.}\ }\textbf
  {\bibinfo {volume} {17}},\ \bibinfo {pages} {161} (\bibinfo {year}
  {1996})}\BibitemShut {NoStop}%
\bibitem [{\citenamefont {S{\o}rensen}\ \emph {et~al.}(2012)\citenamefont
  {S{\o}rensen}, \citenamefont {Lathauwer}, \citenamefont {Icart},\ and\
  \citenamefont {Deneire}}]{Sørensen2012617}%
  \BibitemOpen
  \bibfield  {author} {\bibinfo {author} {\bibfnamefont {M.}~\bibnamefont
  {S{\o}rensen}}, \bibinfo {author} {\bibfnamefont {L.~D.}\ \bibnamefont
  {Lathauwer}}, \bibinfo {author} {\bibfnamefont {S.}~\bibnamefont {Icart}}, \
  and\ \bibinfo {author} {\bibfnamefont {L.}~\bibnamefont {Deneire}},\ }\href
  {\doibase 10.1016/j.sigpro.2011.07.005} {\bibfield  {journal} {\bibinfo
  {journal} {Signal Process.}\ }\textbf {\bibinfo {volume} {92}},\ \bibinfo
  {pages} {617 } (\bibinfo {year} {2012})}\BibitemShut {NoStop}%
\bibitem [{\citenamefont {Xiang}\ \emph {et~al.}(2013)\citenamefont {Xiang},
  \citenamefont {Huang}, \citenamefont {Kan}, \citenamefont {Wei},\ and\
  \citenamefont {Gong}}]{PhysRevLett.110.118702}%
  \BibitemOpen
  \bibfield  {author} {\bibinfo {author} {\bibfnamefont {H.~J.}\ \bibnamefont
  {Xiang}}, \bibinfo {author} {\bibfnamefont {B.}~\bibnamefont {Huang}},
  \bibinfo {author} {\bibfnamefont {E.}~\bibnamefont {Kan}}, \bibinfo {author}
  {\bibfnamefont {S.-H.}\ \bibnamefont {Wei}}, \ and\ \bibinfo {author}
  {\bibfnamefont {X.~G.}\ \bibnamefont {Gong}},\ }\href {\doibase
  10.1103/PhysRevLett.110.118702} {\bibfield  {journal} {\bibinfo  {journal}
  {Phys. Rev. Lett.}\ }\textbf {\bibinfo {volume} {110}},\ \bibinfo {pages}
  {118702} (\bibinfo {year} {2013})}\BibitemShut {NoStop}%
\bibitem [{\citenamefont {Giannozzi}\ \emph {et~al.}(2009)\citenamefont
  {Giannozzi}, \citenamefont {Baroni}, \citenamefont {Bonini}, \citenamefont
  {Calandra}, \citenamefont {Car}, \citenamefont {Cavazzoni}, \citenamefont
  {Ceresoli}, \citenamefont {Chiarotti}, \citenamefont {Cococcioni},
  \citenamefont {Dabo}, \citenamefont {{Dal Corso}}, \citenamefont
  {de~Gironcoli}, \citenamefont {Fabris}, \citenamefont {Fratesi},
  \citenamefont {Gebauer}, \citenamefont {Gerstmann}, \citenamefont
  {Gougoussis}, \citenamefont {Kokalj}, \citenamefont {Lazzeri}, \citenamefont
  {Martin-Samos}, \citenamefont {Marzari}, \citenamefont {Mauri}, \citenamefont
  {Mazzarello}, \citenamefont {Paolini}, \citenamefont {Pasquarello},
  \citenamefont {Paulatto}, \citenamefont {Sbraccia}, \citenamefont {Scandolo},
  \citenamefont {Sclauzero}, \citenamefont {Seitsonen}, \citenamefont
  {Smogunov}, \citenamefont {Umari},\ and\ \citenamefont
  {Wentzcovitch}}]{QE-2009}%
  \BibitemOpen
  \bibfield  {author} {\bibinfo {author} {\bibfnamefont {P.}~\bibnamefont
  {Giannozzi}}, \bibinfo {author} {\bibfnamefont {S.}~\bibnamefont {Baroni}},
  \bibinfo {author} {\bibfnamefont {N.}~\bibnamefont {Bonini}}, \bibinfo
  {author} {\bibfnamefont {M.}~\bibnamefont {Calandra}}, \bibinfo {author}
  {\bibfnamefont {R.}~\bibnamefont {Car}}, \bibinfo {author} {\bibfnamefont
  {C.}~\bibnamefont {Cavazzoni}}, \bibinfo {author} {\bibfnamefont
  {D.}~\bibnamefont {Ceresoli}}, \bibinfo {author} {\bibfnamefont {G.~L.}\
  \bibnamefont {Chiarotti}}, \bibinfo {author} {\bibfnamefont {M.}~\bibnamefont
  {Cococcioni}}, \bibinfo {author} {\bibfnamefont {I.}~\bibnamefont {Dabo}},
  \bibinfo {author} {\bibfnamefont {A.}~\bibnamefont {{Dal Corso}}}, \bibinfo
  {author} {\bibfnamefont {S.}~\bibnamefont {de~Gironcoli}}, \bibinfo {author}
  {\bibfnamefont {S.}~\bibnamefont {Fabris}}, \bibinfo {author} {\bibfnamefont
  {G.}~\bibnamefont {Fratesi}}, \bibinfo {author} {\bibfnamefont
  {R.}~\bibnamefont {Gebauer}}, \bibinfo {author} {\bibfnamefont
  {U.}~\bibnamefont {Gerstmann}}, \bibinfo {author} {\bibfnamefont
  {C.}~\bibnamefont {Gougoussis}}, \bibinfo {author} {\bibfnamefont
  {A.}~\bibnamefont {Kokalj}}, \bibinfo {author} {\bibfnamefont
  {M.}~\bibnamefont {Lazzeri}}, \bibinfo {author} {\bibfnamefont
  {L.}~\bibnamefont {Martin-Samos}}, \bibinfo {author} {\bibfnamefont
  {N.}~\bibnamefont {Marzari}}, \bibinfo {author} {\bibfnamefont
  {F.}~\bibnamefont {Mauri}}, \bibinfo {author} {\bibfnamefont
  {R.}~\bibnamefont {Mazzarello}}, \bibinfo {author} {\bibfnamefont
  {S.}~\bibnamefont {Paolini}}, \bibinfo {author} {\bibfnamefont
  {A.}~\bibnamefont {Pasquarello}}, \bibinfo {author} {\bibfnamefont
  {L.}~\bibnamefont {Paulatto}}, \bibinfo {author} {\bibfnamefont
  {C.}~\bibnamefont {Sbraccia}}, \bibinfo {author} {\bibfnamefont
  {S.}~\bibnamefont {Scandolo}}, \bibinfo {author} {\bibfnamefont
  {G.}~\bibnamefont {Sclauzero}}, \bibinfo {author} {\bibfnamefont {A.~P.}\
  \bibnamefont {Seitsonen}}, \bibinfo {author} {\bibfnamefont {A.}~\bibnamefont
  {Smogunov}}, \bibinfo {author} {\bibfnamefont {P.}~\bibnamefont {Umari}}, \
  and\ \bibinfo {author} {\bibfnamefont {R.~M.}\ \bibnamefont {Wentzcovitch}},\
  }\href {http://www.quantum-espresso.org} {\bibfield  {journal} {\bibinfo
  {journal} {J. Phys.: Condens. Matter}\ }\textbf {\bibinfo {volume} {21}},\
  \bibinfo {pages} {395502} (\bibinfo {year} {2009})}\BibitemShut {NoStop}%
\bibitem [{\citenamefont {Vanderbilt}(1990)}]{PhysRevB.41.7892}%
  \BibitemOpen
  \bibfield  {author} {\bibinfo {author} {\bibfnamefont {D.}~\bibnamefont
  {Vanderbilt}},\ }\href {\doibase 10.1103/PhysRevB.41.7892} {\bibfield
  {journal} {\bibinfo  {journal} {Phys. Rev. B}\ }\textbf {\bibinfo {volume}
  {41}},\ \bibinfo {pages} {7892} (\bibinfo {year} {1990})}\BibitemShut
  {NoStop}%
\bibitem [{\citenamefont {Garrity}\ \emph {et~al.}(2014)\citenamefont
  {Garrity}, \citenamefont {Bennett}, \citenamefont {Rabe},\ and\ \citenamefont
  {Vanderbilt}}]{Garrity2014446}%
  \BibitemOpen
  \bibfield  {author} {\bibinfo {author} {\bibfnamefont {K.~F.}\ \bibnamefont
  {Garrity}}, \bibinfo {author} {\bibfnamefont {J.~W.}\ \bibnamefont
  {Bennett}}, \bibinfo {author} {\bibfnamefont {K.~M.}\ \bibnamefont {Rabe}}, \
  and\ \bibinfo {author} {\bibfnamefont {D.}~\bibnamefont {Vanderbilt}},\
  }\href {\doibase 10.1016/j.commatsci.2013.08.053} {\bibfield  {journal}
  {\bibinfo  {journal} {Comput. Mater. Sci.}\ }\textbf {\bibinfo {volume}
  {81}},\ \bibinfo {pages} {446 } (\bibinfo {year} {2014})}\BibitemShut
  {NoStop}%
\bibitem [{\citenamefont {Elemans}\ \emph {et~al.}(1971)\citenamefont
  {Elemans}, \citenamefont {Laar}, \citenamefont {Veen},\ and\ \citenamefont
  {Loopstra}}]{Elemans1971238}%
  \BibitemOpen
  \bibfield  {author} {\bibinfo {author} {\bibfnamefont {J.~B.}\ \bibnamefont
  {Elemans}}, \bibinfo {author} {\bibfnamefont {B.~V.}\ \bibnamefont {Laar}},
  \bibinfo {author} {\bibfnamefont {K.~V.~D.}\ \bibnamefont {Veen}}, \ and\
  \bibinfo {author} {\bibfnamefont {B.}~\bibnamefont {Loopstra}},\ }\href
  {\doibase 10.1016/0022-4596(71)90034-X} {\bibfield  {journal} {\bibinfo
  {journal} {J. Solid State Chem.}\ }\textbf {\bibinfo {volume} {3}},\ \bibinfo
  {pages} {238 } (\bibinfo {year} {1971})}\BibitemShut {NoStop}%
\bibitem [{\citenamefont {Golub}\ and\ \citenamefont
  {Van~Loan}(1996)}]{Golub:1996:MC:248979}%
  \BibitemOpen
  \bibfield  {author} {\bibinfo {author} {\bibfnamefont {G.~H.}\ \bibnamefont
  {Golub}}\ and\ \bibinfo {author} {\bibfnamefont {C.~F.}\ \bibnamefont
  {Van~Loan}},\ }\href@noop {} {\emph {\bibinfo {title} {Matrix
  Computations}}},\ \bibinfo {edition} {3rd}\ ed.\ (\bibinfo  {publisher}
  {Johns Hopkins University Press},\ \bibinfo {address} {Baltimore, MD},\
  \bibinfo {year} {1996})\BibitemShut {NoStop}%
\bibitem [{\citenamefont {Gander}\ \emph {et~al.}(1989)\citenamefont {Gander},
  \citenamefont {Golub},\ and\ \citenamefont {von Matt}}]{Gander1989815}%
  \BibitemOpen
  \bibfield  {author} {\bibinfo {author} {\bibfnamefont {W.}~\bibnamefont
  {Gander}}, \bibinfo {author} {\bibfnamefont {G.~H.}\ \bibnamefont {Golub}}, \
  and\ \bibinfo {author} {\bibfnamefont {U.}~\bibnamefont {von Matt}},\ }\href
  {\doibase 10.1016/0024-3795(89)90494-1} {\bibfield  {journal} {\bibinfo
  {journal} {Linear Alg. Appl.}\ }\textbf {\bibinfo {volume} {114-115}},\
  \bibinfo {pages} {815 } (\bibinfo {year} {1989})}\BibitemShut {NoStop}%
\end{thebibliography}%
\bibliographystyle{apsrev4-1}

\end{document}